\expandafter\ifx\csname LaTeX\endcsname\relax
      \let\maybe\relax
\else \immediate\write0{}
      \message{You need to run TeX for this, not LaTeX}
      \immediate\write0{}
      \makeatletter\let\maybe\@@end
\fi
\maybe

\magnification=\magstephalf

\hsize=5.25truein
\vsize=8.3truein
\hoffset=0.37truein

\newdimen\frontindent \frontindent=.45truein
\newdimen\theparindent \theparindent=20pt


\let\em=\it

\font\tencsc=cmcsc10
\font\twelvebf=cmbx10 scaled 1200
\font\bmit=cmmib10  \font\twelvebmit=cmmib10 scaled 1200
\font\sixrm=cmr6 \font\sixi=cmmi6 \font\sixit=cmti8 at 6pt

\font\eightrm=cmr8  \let\smallrm=\eightrm
\font\eighti=cmmi8  \let\smalli=\eighti
\skewchar\eighti='177
\font\eightsy=cmsy8
\skewchar\eightsy='60
\font\eightit=cmti8
\font\eightsl=cmsl8
\font\eightbf=cmbx8
\font\eighttt=cmtt8
\def\eightpoint{\textfont0=\eightrm \scriptfont0=\fiverm 
                \def\rm{\fam0\eightrm}\relax
                \textfont1=\eighti \scriptfont1=\fivei 
                \def\mit{\fam1}\def\oldstyle{\fam1\eighti}\relax
                \textfont2=\eightsy \scriptfont2=\fivesy 
                \def\cal{\fam2}\relax
                \textfont3=\tenex \scriptfont3=\tenex 
                \def\it{\fam\itfam\eightit}\let\em=\it
                \textfont\itfam=\eightit
                \def\sl{\fam\slfam\eightsl}\relax
                \textfont\slfam=\eightsl
                \def\bf{\fam\bffam\eightbf}\relax
                \textfont\bffam=\eightbf \scriptfont\bffam=\fivebf
                \def\tt{\fam\ttfam\eighttt}\relax
                \textfont\ttfam=\eighttt
                \setbox\strutbox=\hbox{\vrule
                     height7pt depth2pt width0pt}\baselineskip=9pt
                \let\smallrm=\sixrm \let\smalli=\sixi
                \rm}


\catcode`@=11
 
\def\vfootnote#1{\insert\footins\bgroup\eightpoint
     \interlinepenalty=\interfootnotelinepenalty
     \splittopskip=\ht\strutbox \splitmaxdepth=\dp\strutbox
     \floatingpenalty=20000
     \leftskip=0pt \rightskip=0pt \parskip=1pt \spaceskip=0pt \xspaceskip=0pt
     \everydisplay={}
     \smallskip\textindent{#1}\footstrut\futurelet\next\fo@t}
 
\newcount\notenum

\def\note{\global\advance\notenum by 1
    \edef\n@tenum{$^{\the\notenum}$}\let\@sf=\empty
    \ifhmode\edef\@sf{\spacefactor=\the\spacefactor}\/\fi
    \n@tenum\@sf\vfootnote{\n@tenum}}


\tabskip1em

\newtoks\pream \pream={#\strut}
\newtoks\lpream \lpream={&#\hfil}
\newtoks\rpream \rpream={&\hfil#}
\newtoks\cpream \cpream={&\hfil#\hfil}
\newtoks\mpream \mpream={&&\hfil#\hfil}

\newcount\ncol \def\ncolp{\advance\ncol by 1}
\def\atalias#1{
    \ifx#1l\edef\xpream{\pream={\the\pream\the\lpream}}\xpream\ncolp\fi
    \ifx#1r\edef\xpream{\pream={\the\pream\the\rpream}}\xpream\ncolp\fi
    \ifx#1c\edef\xpream{\pream={\the\pream\the\cpream}}\xpream\ncolp\fi}
\catcode`@=\active

\def\taborl#1{\omit\unskip#1\hfil}
\def\taborc#1{\omit\hfil#1\hfil}
\def\taborr#1{\omit\hfil#1}
\def\multicol#1{\multispan#1\let\omit\relax}

\def\table#1\par{\midinsert\offinterlineskip\everydisplay{}
    \let@\atalias \let\l\taborl \let\r\taborr \let\c\taborc
    \def\space{\noalign{\vskip2pt}}
    \def\tablerule{\omit&\multispan{\the\ncol}\hrulefill\cr}
    \def\onerule{\space\space\tablerule\space\space}
    \def\tworules{\space\space\tablerule\space\tablerule\space\space}
    \def\annot##1\\{&\multispan{\the\ncol}##1\hfil\cr}
    \def\\{\let\\=\cr
           \edef\xpream{\pream={\the\pream\the\mpream}}\xpream
           \edef\starthalign{$$\vbox\bgroup\halign\bgroup\the\pream\cr}
           \starthalign
           \annot\hfil\tencsc Table #1\\ \noalign{\medskip}}
    \let\par\endtable}

\edef\endtable{\noalign{\vskip-\bigskipamount}\egroup\egroup$$\endinsert}

\let\plainmidinsert=\midinsert
\def\eightpttable{\def\midinsert{\let\midinsert=\plainmidinsert
    \plainmidinsert\eightpoint\tabskip 1em}\table}



\newif\iftitlepage

\def\raggedright{\rightskip 0pt plus .2\hsize\relax}

\let\caret=^ \catcode`\^=13 \def^#1{\ifmmode\caret{#1}\else$\caret{#1}$\fi}

\def\title#1\par{\vfill\supereject\begingroup
                 \global\titlepagetrue
                 \leftskip=\frontindent\parindent=0pt\parskip=0pt
                 \frenchspacing \eqnum=0
                 \gdef\runningtitle{#1}
                 \null\vskip-22.5pt\copy\volbox\vskip18pt
                 {\titlestyle#1\bigskip}}
\def\titlestyle{\raggedright\bf\twelvebf\textfont1=\twelvebmit
                \let\smallrm=\tenbf \let\smalli=\bmit
                \baselineskip=1.2\baselineskip}
\def\shorttitle#1\par{\gdef\runningtitle{#1}}
\def\author#1\par{{\raggedright#1\medskip}}

\def\shortauthor#1\par{\gdef\runningauthors{#1}}

\def\affil#1\par{{\raggedright\it#1\smallskip}}
\def@#1{\ifhmode\qquad\fi\leavevmode\llap{^{#1}}\ignorespaces}
\def\abstract{\smallskip\medskip{\bf Abstract: }}

\def\maybebreak#1{\vskip0pt plus #1\hsize \penalty-500
                  \vskip0pt plus -#1\hsize}

\def\maintextmode{\leftskip=0pt\parindent=\theparindent
                  \parskip=\smallskipamount\nonfrenchspacing}

\def\maintext#1\par{\bigskip\medskip\maintextmode\noindent}

\newcount\secnum
\def\section#1\par{\ifnum\secnum=0\medskip\maintextmode\fi
    \advance\secnum by 1 \bigskip\maybebreak{.1}
    \subsecnum=0
    \hang\noindent\hbox to \parindent{\bf\the\secnum.\hfil}{\bf#1}
    \smallskip\noindent}

\newcount\subsecnum
\def\subsection#1\par{\ifnum\subsecnum>0\medskip\maybebreak{.1}\fi
    \advance\subsecnum by 1
    \hang\noindent\hbox to \parindent
       {\it\the\secnum.\the\subsecnum\hfil}{\it#1}
    \par\noindent}

\def\references\par{\bigskip\maybebreak{.1}\parindent=0pt
    \everypar{\hangindent\theparindent\hangafter1}
    \leftline{\bf References}\smallskip}

\def\appendix#1\par{\bigskip\maybebreak{.1}\maintextmode
    \advance\secnum by 1 \bigskip\maybebreak{.1}
    \leftline{\bf Appendix: #1}\smallskip\noindent}

\def\acknowl{\medskip\noindent}

\def\bye{\endgroup\vfill\supereject\end}


\newbox\volbox
\setbox\volbox=\vbox{\hsize=.5\hsize \raggedright
       \sixit\baselineskip=7.2pt \noindent
       Twelfth Annual Florida Workshop in 
       Nonlinear Astronomy and Physics,
       Long Range Correlations in Astrophysical
       and Other Systems.
       To appear in Annals of the New York Academy of Sciences,
       Eds.\ J.R.~Buchler, J.~Dufty \& H.~Kandrup}


\input epsf

\def\figureps[#1,#2]#3.{\midinsert\parindent=0pt\eightpoint
    \vbox{\epsfxsize=#2\centerline{\epsfbox{#1}}}
    \def\par{\endgraf\endinsert}{\bf Figure#3.}}

\def\figuretwops[#1,#2,#3]#4.{\midinsert\parindent=0pt\eightpoint
    \vbox{\centerline{\epsfxsize=#3\epsfbox{#1}\hfil
                      \epsfxsize=#3\epsfbox{#2}}}
     \def\par{\endgraf\endinsert}{\bf Figure#4.}}

\def\figurespace[#1]#2.{\midinsert\parindent=0pt\eightpoint
    \vbox to #1 {\vfil\centerline{\tenit Stick Figure#2 here!}\vfil}
    \def\par{\endgraf\endinsert}{\bf Figure#2.}}


\headline={\iftitlepage\hfil\else
              \ifodd\pageno\hfil\tensl\runningtitle
                    \kern1pc\tenbf\folio
               \else\tenbf\folio\kern1pc
                    \tensl\runningauthors\hfil\fi
           \fi}
\footline{\iftitlepage\tenbf\hfil\folio\hfil\else\hfil\fi}
\output={\plainoutput\global\titlepagefalse}


\newcount\eqnum
\everydisplay{\puteqnum}  
\def\puteqnum#1$${#1\global\advance\eqnum by 1\eqno(\the\eqnum)$$}
\def\namethiseqn#1{\xdef#1{\the\eqnum}}

 
\newcount\mpageno
\mpageno=\pageno  \advance\mpageno by 1000
 
\def\advancepageno{\global\advance\pageno by 1
                   \global\advance\mpageno by 1 }

\openout15=inx
\def\index#1{\write15{{#1}{\the\mpageno}}\ignorespaces}


\def\LaTeX{{\rm L\kern-.36em\raise.3ex\hbox{\tencsc a}\kern-.15em
    T\kern-.1667em\lower.7ex\hbox{E}\kern-.125emX}}

\def\[#1]{\raise.2ex\hbox{[}#1\raise.2ex\hbox{]}}

\def\witchbox#1#2#3{\hbox{$\mathchar"#1#2#3$}}
\def\leqsim{\mathrel{\rlap{\lower3pt\witchbox218}\raise2pt\witchbox13C}}
\def\geqsim{\mathrel{\rlap{\lower3pt\witchbox218}\raise2pt\witchbox13E}}

\def\<#1>{\langle#1\rangle}


{\obeyspaces\gdef {\ }}

\catcode`@=12 \let\@=@ \catcode`@=13
\def\+{\catcode`\\=12\catcode`\$=12\catcode`\&=12\catcode`\#=12%
       \catcode`\^=12\catcode`\_=12\catcode`\~=12\catcode`\%=12%
       \catcode`\@=0\tt}
\def\({\endgraf\bgroup\let\par=\endgraf\parskip=0pt\vskip3pt
       \eightpoint \def\/{{\eightpoint$\langle$Blank line$\rangle$}}
       \catcode`\{=12\catcode`\}=12\+\obeylines\obeyspaces}
\def\){\vskip1pt\egroup\vskip-\parskip\noindent\ignorespaces}


\def\ltsima{$\; \buildrel < \over \sim \;$}
\def\simlt{\lower.5ex\hbox{\ltsima}}
\def\gtsima{$\; \buildrel > \over \sim \;$}
\def\simgt{\lower.5ex\hbox{\gtsima}}

\title SELF-CONSISTENT GRAVITATIONAL CHAOS

\shorttitle Gravitational Chaos

\author David Merritt and Monica Valluri

\shortauthor Merritt \& Valluri

\affil Rutgers University, New Brunswick, NJ

\abstract 
The motion of stars in the gravitational potential of a triaxial
galaxy is generically chaotic.
However, the timescale over which the chaos manifests itself in the
orbital motion is a strong function of the degree of central
concentration of the galaxy.
Here, chaotic diffusion rates are presented for orbits in triaxial
models with a range of central density slopes and nuclear
black-hole masses.
Typical diffusion times are found to be less than a galaxy lifetime
in triaxial models where the density increases more rapidly than
$\sim r^{-1}$ at the center, or which contain black holes with masses
that exceed $\sim 0.1\%$ of the galaxy mass.
When the mass of a central black hole exceeds roughly $0.02\
M_{gal}$, there is a transition to global stochasticity and the
galaxy evolves to an axisymmetric shape in little more than a
crossing time.
This rapid evolution may provide a negative feedback mechanism that
limits the mass of nuclear black holes to a few percent of the
stellar mass of a galaxy.

\section Introduction

The gravitational force that determines the motion of a star in a 
galaxy has two components: the smooth force generated by
the overall mass distribution, and the non-smooth force that 
results from close encounters between individual stars.
The relative importance of the two components after one
orbital period is roughly equal to $0.1N/\log N$, where 
$N$ is the number of stars in the galaxy.$^1$  
For a typical galaxy with $N\approx 10^{11}$ stars, close encounters 
between stars are unimportant and the dynamics are 
essentially collisionless, at least over timescales of $10^2-10^3$ 
orbital periods that correspond to galaxy lifetimes.

Motion in a smooth gravitational field becomes quite simple if the 
number of isolating integrals equals or exceeds the number of degrees of 
freedom, and much work in galactic dynamics has focussed on 
finding integrable or near-integrable models for galactic potentials.$^2$
Kuzmin$^{3,4}$ showed that there is a unique, 
ellipsoidally-stratified mass model for which the corresponding 
potential has three global integrals of the motion, quadratic
in the velocities.
Kuzmin's model -- explored in detail by de Zeeuw,$^5$ who christened 
it the ``Perfect Ellipsoid'' -- has a large, 
constant-density core in which the orbital motion is 
that of a 3-D harmonic oscillator.
Every orbit in the core of the Perfect Ellipsoid fills
a rectangular parallelepiped, or ``box.''
At higher energies in the Perfect Ellipsoid, box orbits persist, 
and three new orbit families appear: the ``tubes,'' orbits that 
preserve the direction of their motion around either the long or 
short axis of the figure.
Tube orbits respect an integral of the motion analogous to the 
angular momentum, and hence -- unlike box orbits -- avoid the center.
When two of the axis lengths of an ellipsoidal model are equal, the 
box orbits disappear, and all trajectories belong to a single family 
of tube orbits that circulate about the axis of symmetry.
Thus, box orbits are uniquely associated with the triaxial 
geometry.

Unfortunately, the Perfect Ellipsoid does not look very much like 
real elliptical galaxies.
Its density falls off as $\rho\propto r^{-4}$ at large radii; 
both the luminosity and mass density in real galaxies fall off 
more slowly, as $r^{-3}$ or $r^{-2}$.
And recent observations of galactic nuclei demonstrate that 
constant-density cores do not exist -- the density of 
starlight in real galaxies always rises 
monotonically at small radii, roughly as a power law.$^6$
The density profile in these power-law ``cusps'' rises as steeply 
as $\rho\propto r^{-2}$ in faint elliptical galaxies, while the cusps 
in brighter galaxies are typically shallower.$^7$
There is also increasingly strong evidence that many elliptical 
galaxies and bulges contain central massive objects,
possibly the black holes that are thought to have powered quasars.$^8$
While the masses of these dark central components are often very 
uncertain, typical estimates are $10^{-3}\simlt M_{BH}/M_{gal}\simlt
10^{-2}$, where $M_{BH}$ is the black hole mass inferred 
from the orbital motions of surrounding stars and gas, and $M_{gal}$ is 
the stellar mass of the host galaxy or (in the case of a spiral 
galaxy) the mass of the stellar bulge.
Some galaxies, like M32, the dwarf companion to the nearby 
Andromeda galaxy, are known to contain both a steep stellar cusp 
($\rho\propto r^{-1.6})$ and a dynamically-significant black hole 
($M_{BH}/M_{gal}\sim 0.003)$.

The existence of box orbits in the Perfect Ellipsoid is tied to 
the stability of the long-, or $x$-axis orbit.$^9$
The $x$-axis orbit is unstable at most energies to lateral perturbations 
in triaxial models where the density increases more 
rapidly than $\sim r^{-1}$ near the center.$^{10}$
The instability first appears through the bifurcation of a $1:2$ 
resonant orbit, the $x-z$ ``banana'' boxlet.$^{11}$
The $x$-axis orbit is likewise unstable in any triaxial model 
with a central singularity.$^{12}$
It follows that bona-fide box orbits do not exist in the 
majority of triaxial potentials corresponding to real elliptical 
galaxies; in their place, we would expect to 
find either stochastic orbits, or regular orbits associated with 
minor resonances (like the banana) that avoid the center.

The non-existence of box orbits has important consequences for 
the self-consistent dynamics of elliptical galaxies.
Schwarzschild$^{13,14}$ and Statler$^{15}$ found that box 
orbits -- particularly the thin boxes that remain close 
to the long axis -- were crucial for reconstructing the 
distribution of mass in triaxial models.
Their work was based on mass models with smooth cores.
Merritt \& Fridman$^{16}$ attempted to construct 
self-consistent triaxial models with central density cusps, 
after excluding the stochastic boxlike orbits, or  
replacing them with invariant ensembles representing a uniform 
population of stochastic phase space.
Completely stationary solutions could not be found; only
quasi-equilibrium solutions, in which stochastic phase space was 
populated in a non-uniform way, could successfully 
reproduce the density at all points in the model.
An extension of this work to triaxial models with a range 
of axis ratios$^{17}$ revealed that stationary models with $r^{-2}$ cusps 
could only be constructed if the figure was nearly axisymmetric.

One important question left unanswered by these equilibrium studies is 
the timescale over which chaos implies changes in the 
self-consistent structure of a galaxy.
Here, results from two recent studies that address this 
question$^{18, 19}$ are presented.

\section Diffusion of Stochastic Orbits

In order for chaos to be relevant to real galaxies, it must 
produce a significant change in the region visited by an orbit 
after just a few tens or hundreds of orbital periods -- the approximate 
lifetime of a galaxy.
In a pioneering study, Goodman \& Schwarzschild$^9$ found that 
the boxlike orbits in a triaxial model with a smooth core were 
often stochastic, but that the orbital motion was essentially
regular for at least $50$ oscillations.
On the other hand, Merritt \& Fridman$^{16}$ found that the 
stochasticity in 
triaxial models with $\rho\propto r^{-2}$ density cusps produced 
significant changes in the appearance of boxlike orbits after just a few 
tens of oscillations.
Merritt \& Valluri$^{20}$ went on to calculate timescales for 
mixing in these strongly chaotic potentials; they found that 
ensembles of stochastic trajectories evolved toward invariant 
distributions -- corresponding to an approximately uniform filling 
of stochastic phase space -- on timescales of only $10^1-10^2$ 
orbital periods.
Taken together, these results suggest that the characteristic time 
over which chaos manifests itself in the orbital motion is strongly
dependent on the central concentration of a triaxial model.
Presumably, this dependence reflects the sensitivity of boxlike orbits
to deflections that occur during close passages to the galaxy center.

A useful way to quantify the diffusion of stochastic orbits 
has been described by Laskar.$^{21,22}$
His ``frequency mapping'' technique is based on the fact that 
a regular orbit can be characterized by the three fundamental 
frequencies $\omega_{i,\ i=1,2,3}$ that define motion on an 
invariant torus.
These frequencies can be computed with high accuracy by 
integrating an orbit for a finite length of time $\Delta T$, 
storing its phase-space coordinates at regular intervals, and 
decomposing the motion into Fourier components.
Carrying out this procedure for a large number of orbits at a given 
energy and plotting $\omega_1/\omega_3$ vs. $\omega_2/\omega_3$ 
gives the ``frequency map,'' a regular grid of points in 
frequency space.
The same procedure can be carried out for stochastic orbits, but now 
the values of the $\omega_i$ will depend on the integration 
interval -- a stochastic orbit is not confined to a single 
invariant torus and its ``characteristic frequencies'' will change 
with time as it diffuses along the energy surface.
A strongly chaotic orbit will migrate through frequency space 
in just a few orbital periods, while a weakly chaotic orbit 
can remain close to a single invariant torus for a very long time.
The diffusion is apparent as a distortion of the frequency map; 
the size of the distortion is proportional to the distance that 
the orbits have moved in frequency space during the 
period of integration $\Delta T$.

Papaphilippou \& Laskar$^{23,24}$ demonstrated the usefulness 
of the frequency mapping technique for galaxy dynamics by 
using it to study orbital motion in the logarithmic potential,
$$
\Phi(x,y,z) = \log\left(m^2+ r_c^2\right),\ \ \ \  
m^2 = {x^2\over a^2} + {y^2\over b^2} + {z^2\over c^2}. 
$$
Equation (1) is the gravitational potential generated by a 
triaxial mass model in which the density falls off as $r^{-2}$ 
outside of the core radius $r_c$.
Papaphilippou \& Laskar established that the motion of boxlike orbits 
in the logarithmic potential is generically chaotic, 
and showed that much of the chaos is associated with motion out 
of the principal planes. 

A mass model that better represents the distribution of starlight 
near the centers of elliptical galaxies and bulges is 
Dehnen's law,$^{25}$
$$
\rho(m) = \rho_0 m^{-\gamma}(1+m)^{-(4-\gamma)} .
$$
Dehnen's formula allows the steepness of the central density cusp to be 
adjusted through the parameter $\gamma$ -- a useful feature, 
since real elliptical galaxies and bulges have cusps with a range of 
slopes, $0\simlt\gamma\simlt 2$.
Figure 1 shows frequency maps of boxlike orbits in triaxial 
potentials derived from Dehnen's density law, with $c/a=1/2$ and
$b/a=0.79$, and five different values of $\gamma$.
All the orbits in a given map have the same energy, corresponding 
roughly to the gravitational potential at the half-mass radius.
The fundamental frequencies were computed over a time interval 
$\Delta T$ equal to 50 periods of the $x$-axis orbit -- roughly 
a galaxy lifetime.
Approximately $10^4$ orbits were integrated for each map; the 
initial conditions consisted of a grid of points distributed
uniformly over the equipotential surface.$^{18}$

The frequency maps of Figure 1 show a clear transition from 
nearly regular behavior over $50$ orbital periods when $\gamma = 0$ 
or $0.5$, to clearly chaotic motion when $\gamma = 1.5$ or $2$.
Many of the orbits can be seen to lie in resonance zones, regions 
associated with rational resonant layers $l\omega_1 + m\omega_2 + 
n\omega_3 = 0$, which appear as lines in the frequency maps.
The last resonance to strongly influence the motion as 
$\gamma$ approaches 2 is the $1:2$ $x-z$ banana resonance, 
which generates the vertical lines $\omega_1/\omega_3=0.5$ in Figure 1.
For smaller values of $\gamma$ the motion is influenced by a 
number of different resonances, some of rather high order.

\figureps[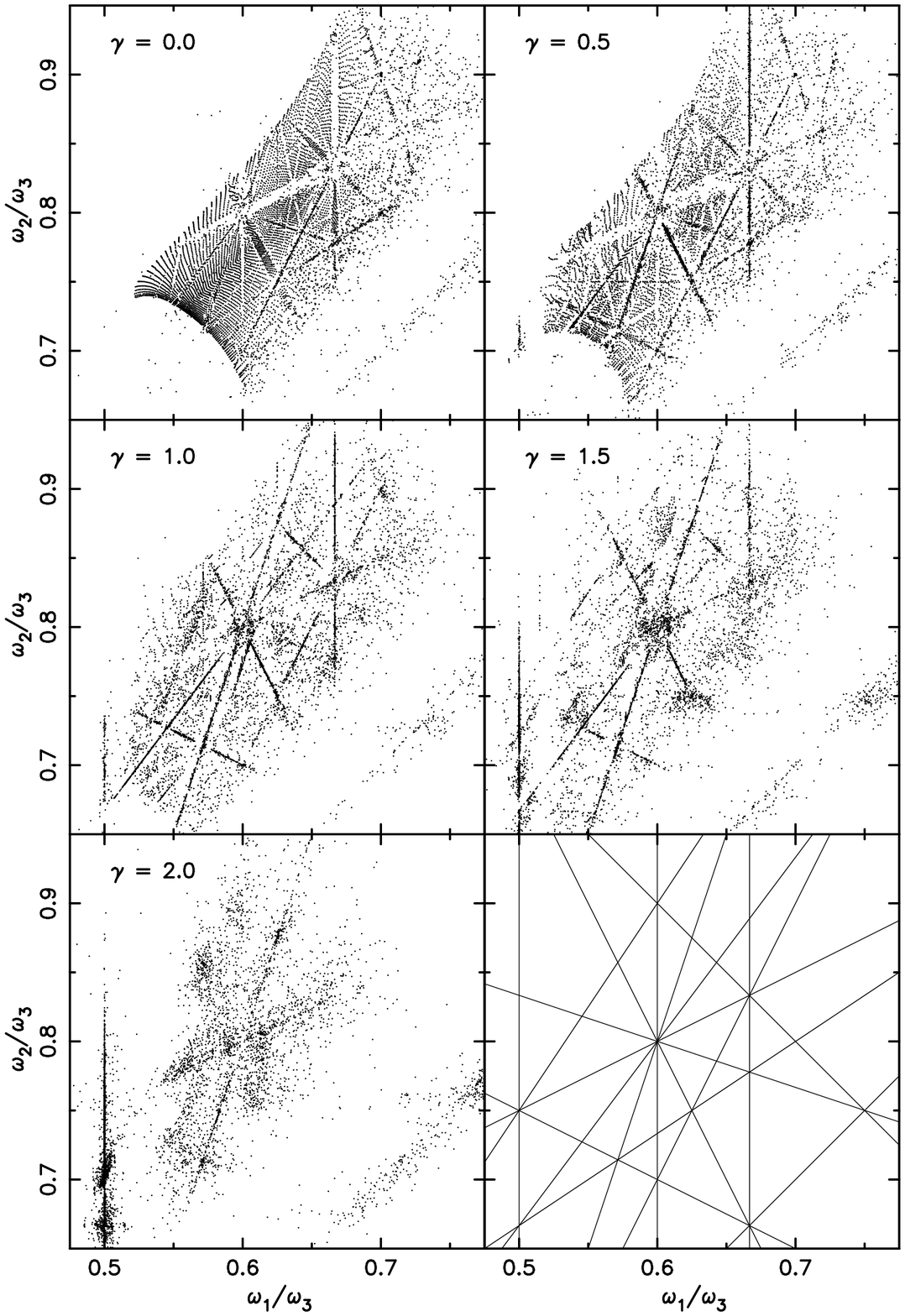,1.\hsize]
1. Frequency maps of boxlike orbits in triaxial 
potentials corresponding to the mass distribution of Eq. (2),
with various cusp slopes $\gamma$.$^{18}$
$\omega_1, \omega_2$ and $\omega_3$ are the fundamental 
frequencies associated with oscillations along the long, 
intermediate and short axes of the triaxial ellipsoid.
The nearly regular grid of points in the upper left-hand panel 
indicates that most of the orbits in the $\gamma=0$ model mimic
regular orbits over 50 oscillations.
Departures from a regular grid imply that diffusion has taken 
place in frequency space, i.e. that the motion is stochastic.
The bottom right panel shows the most important resonance layers, 
defined by $l\omega_1 + m\omega_2 + n\omega_3 = 0$ with integer 
$\{l,m,n\}$.

\figureps[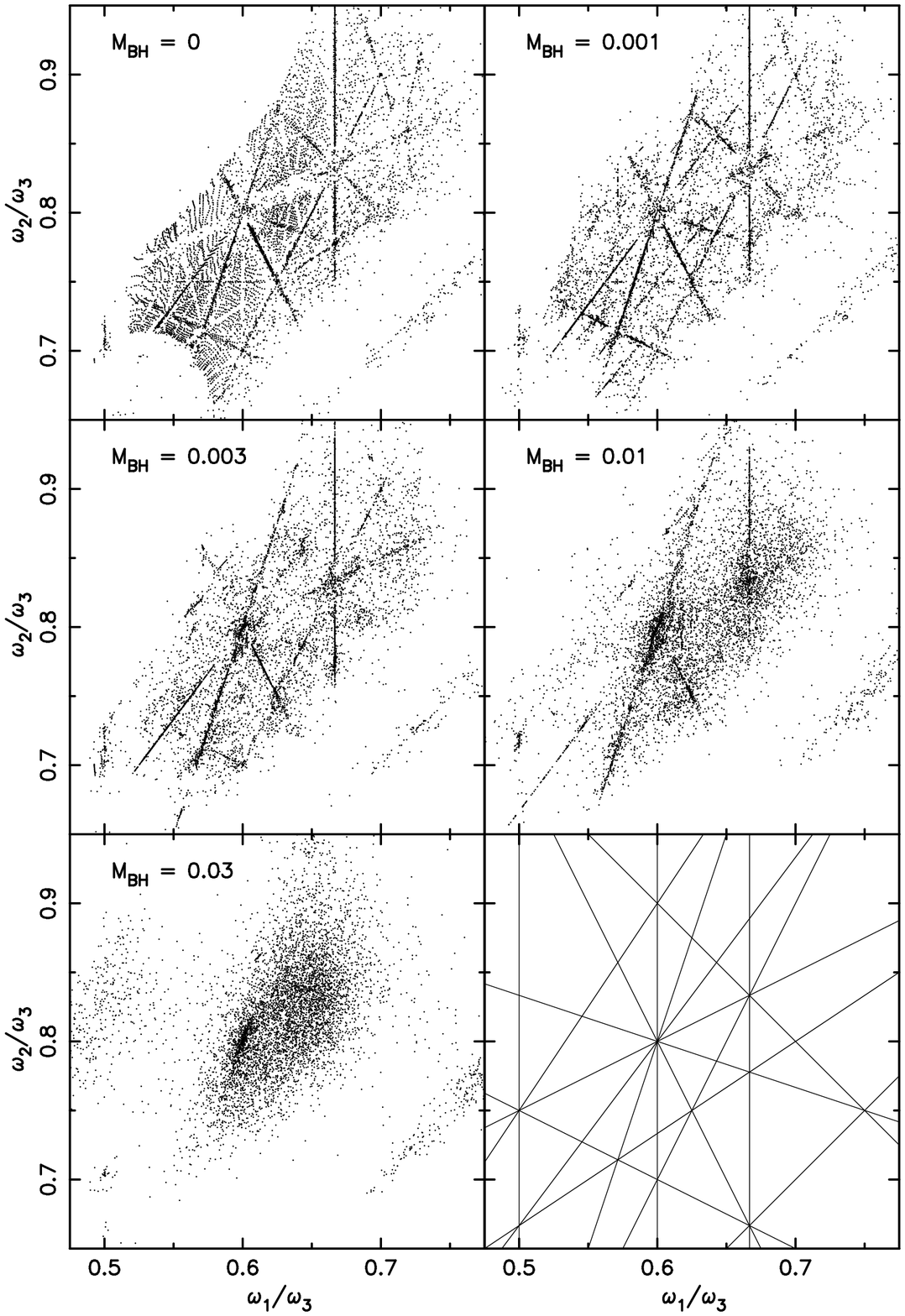,1.\hsize]
2. Like Figure 1, for boxlike orbits in Dehnen's 
model (Eq. 2) with $\gamma=0.5$ and with an added central point 
mass containing various fractions $M_{BH}$ of the total mass of 
the model.$^{18}$

By computing the fundamental frequencies over two 
adjacent time intervals, each of length $\Delta T$, one can
define a rate of diffusion in frequency space $\Delta\omega/\Delta T=
{\rm max}\{|\Delta\omega_i|,_{i=1,2,3}\}/\Delta T$.$^{26}$
The distribution of $\Delta\omega$'s is shown in Figure 3 for each set 
of orbits used in the construction of the frequency maps of Figure 1.
Remarkably, there are no separate peaks associated with regular 
orbits, i.e. orbits with constant $\omega$'s (such orbits would be 
expected to have $\Delta\omega/\omega_0\approx 10^{-4}$, the 
approximate accuracy of the numerical routine that calculates the 
fundamental frequencies.)
Instead, the spectrum of diffusion rates continues to rise toward 
small $\Delta\omega$, roughly as a power law.
While there undoubtedly exist regular boxlike orbits associated with 
stable resonances like the banana, Figure 3 suggests that 
the phase space volume associated with such orbits is very small.
As $\gamma$ approaches zero, the slope of the distribution
increases, i.e. a smaller fraction of the boxlike orbits exhibit 
strong diffusion.
But even for $\gamma=0$ there is no hint of a separate 
population of regular orbits -- rather, the typical diffusion 
times become much longer than $50$ orbital periods, indicating 
that the majority of boxlike orbits are trapped for long periods 
of time in narrow regions of phase space.

\figureps[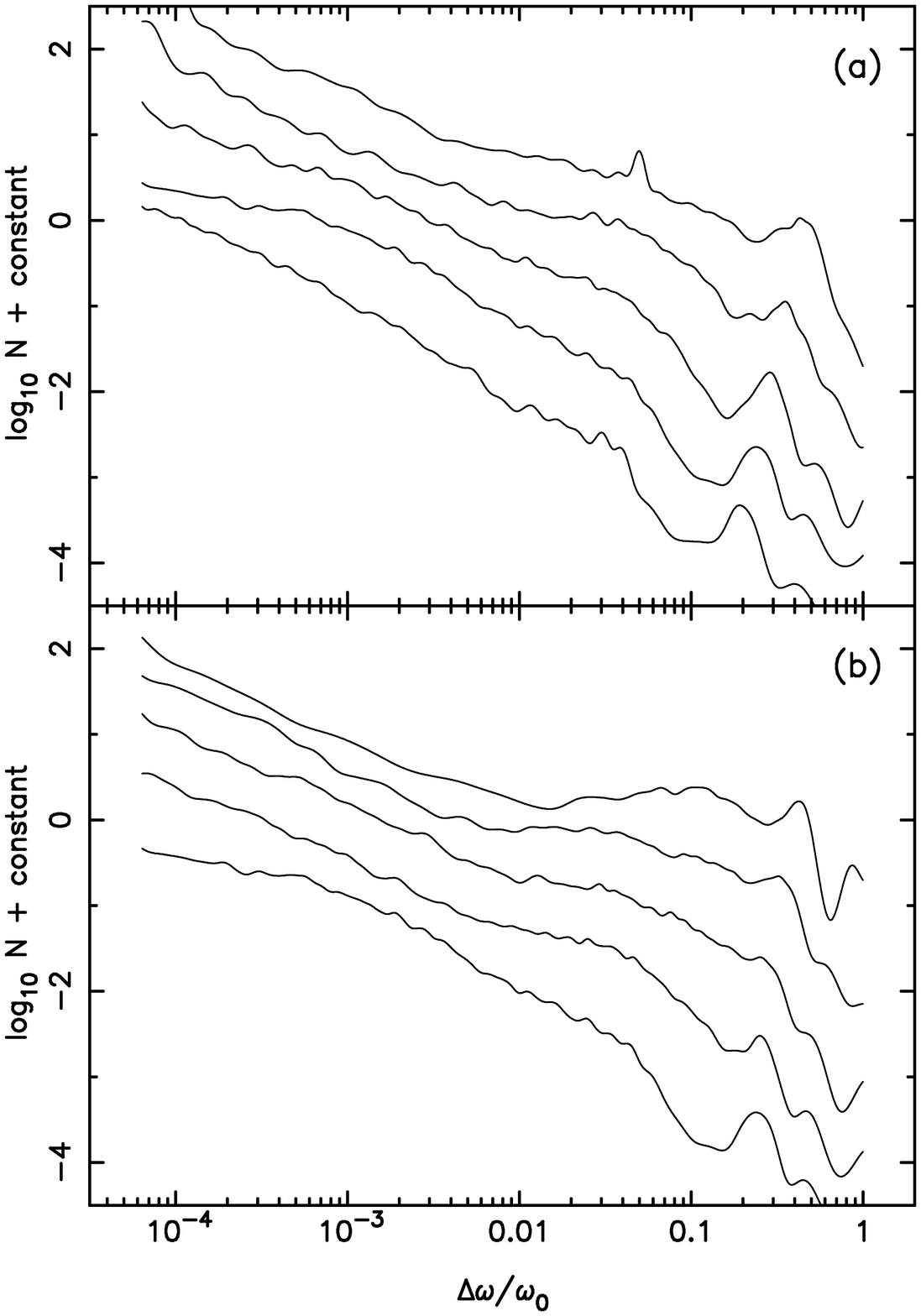,0.7\hsize]
3. Spectra of diffusion rates for the orbits 
plotted in (a) Figure 1 and (b) Figure 2.
The model parameters $\gamma$ and $M_{BH}$ increase upward.
$\Delta\omega$ is the change in fundamental frequencies over 
$50$ orbital periods; $\omega_0$ is the frequency of the 
long-axis orbit.

These results suggest that the motion of boxlike orbits in 
triaxial potentials is generically chaotic, but that the 
distribution of diffusion rates varies systematically 
with the degree of central concentration of the model.
In models with low central concentrations, typical diffusion 
times are long compared to $50$ orbital periods, but as the 
central concentration is increased, a larger fraction of the 
orbits are able to diffuse significantly in a galaxy lifetime.
A very crude index of the importance of the diffusion 
can be defined as the fraction of boxlike orbits for which 
$\Delta\omega/\omega_0>0.1$, i.e. the fraction of orbits which experience 
a 10\% or greater change in their fundamental frequencies over 50 
orbital periods.
Figure 4 shows that this fraction increases from $\sim 10\%$ at 
$\gamma=0$ to $\sim 20\%$ at $\gamma=1$ and $\sim 50\%$ at $\gamma=2$.
Thus, a typical boxlike orbit is essentially regular over the lifetime 
of a galaxy in triaxial potentials with $\gamma\simlt 1$, 
and essentially chaotic when $\gamma\approx 2$.
This conclusion is consistent with the very different 
configuration-space appearance of boxlike orbits integrated in triaxial 
potentials with cores$^9$ and with steep cusps$^{16}$.

\figureps[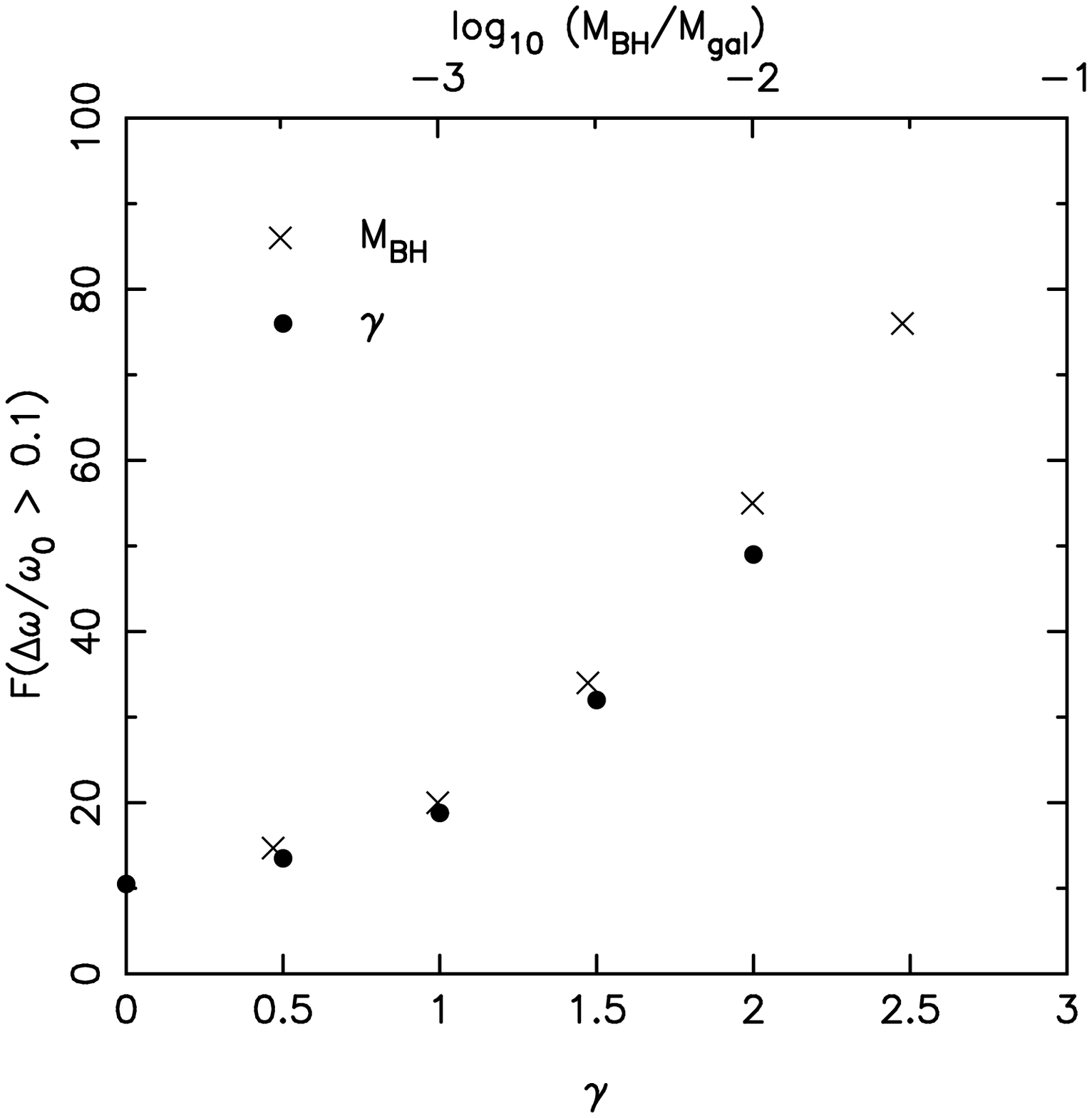,0.6\hsize]
4. Fraction of boxlike orbits that undergo strong 
diffusion, $\Delta\omega/\omega_0\ge 0.1$, in $50$ orbital 
periods.

The effect of a central singularity on the frequency maps is 
shown in Figure 2, computed from Dehnen's $\gamma=0.5$ model with 
an added central point mass.
When the black hole mass $M_{BH}$ is less than about $0.1\%$ of the 
galaxy mass, the frequency map differs only slightly from that of 
the model with $M_{BH}=0$.
As $M_{BH}$ is increased to $1\%$, the motion becomes clearly 
chaotic, and for $M_{BH}=3\%$ there is hardly a trace of 
structure remaining in the frequency map.
Figure 3 verifies that the spectrum of diffusion rates becomes 
very shallow for large $M_{BH}$, even increasing toward large 
$\Delta\omega$ when $M_{BH}\simgt 0.3\%$.
The fraction of boxlike orbits undergoing strong diffusion is 
illustrated in Figure 4; for $M_{BH}=3\%$, fully $3/4$ of the 
orbits evolve strongly over $50$ orbital periods.

\figureps[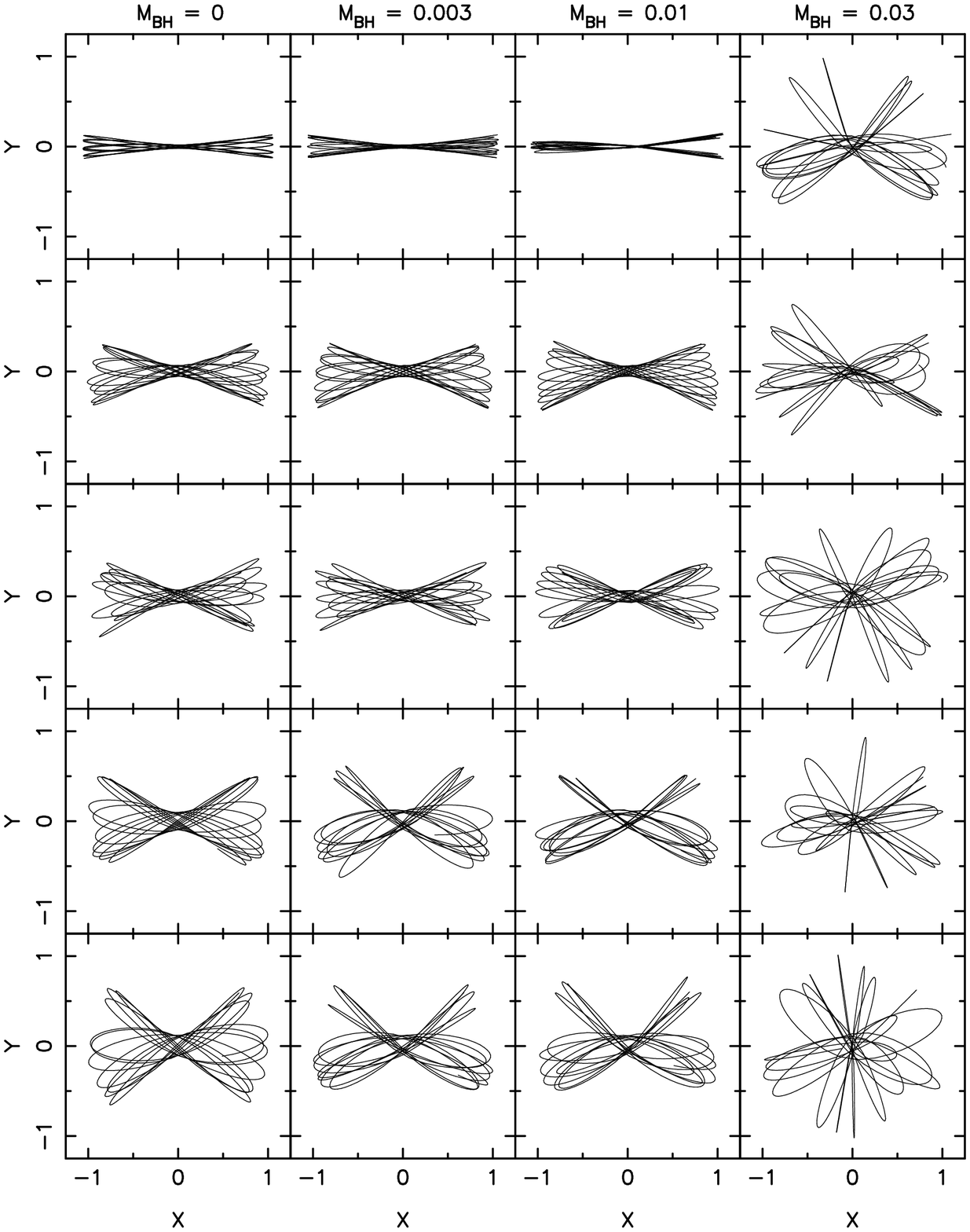,1.15\hsize]
5. Test-particle integrations in a triaxial model 
with a smooth core and a central black hole.
The $z$-axis is the short axis of the figure.
Each of the five orbits is defined by its starting point on the 
equipotential surface, which remained fixed in angular position
as $M_{BH}$ was increased.
Orbits were integrated for approximately 10 full oscillations.
Chaos induces substantial changes in the orbital shapes over this 
period of time when $M_{BH}\simgt 0.02 M_{gal}$.

The frequency maps paint a more complex picture of the
phase space of triaxial potentials than was apparent from earlier 
studies based on cruder techniques for characterizing the stochasticity.
For instance, instability timescales of boxlike orbits inferred from 
Liapunov exponents are only $3-5$ orbital periods in triaxial models 
like those studied here, even in models with smooth cores.$^{20}$
This is much shorter than the typical diffusion time inferred from 
the frequency map analysis.
The apparent discrepancy can be understood by recognizing that 
the motion in weakly chaotic potentials is confined over long 
periods of time to narrow regions in phase space.
The Liapunov exponents measure only the divergence rate of 
nearby trajectories within these limited regions, and not the 
physically more interesting timescale for diffusion from one such 
region to another.
As the degree of central concentration of the model is increased, 
the stochastic layers increase in size until they overlap, and a 
large fraction of the boxlike orbits are able to wander ergodically 
over the energy surface in a finite length of time.
This change in the structure of the phase space is reflected in 
the spectrum of diffusion rates, which becomes shallower 
as $\gamma$ or $M_{BH}$ are increased.

Dynamical systems often exhibit a transition to ``global 
stochasticity'': as a perturbation parameter is increased, 
there is a sudden change from a regime in which the stochastic 
motion is closely bounded by KAM surfaces, to a regime where 
the stochastic motion is interconnected over large portions of 
the phase space.
(Stochastic phase space in a 3 DOF system is always interconnected
through the Arnold web, but Arnold diffusion is extremely 
slow unless the stochastic regions overlap.)
In the globally-stochastic regime, there are few barriers to the 
motion, and stochastic orbits can wander over the full energy 
surface in little more than an orbital period.
Is there a transition to global stochasticity in the triaxial 
models discussed here?
Hints of such a transition can be seen in Figure 4, which shows that 
the fraction of boxlike orbits that evolve strongly in $50$ 
orbital periods becomes large as $M_{BH}$ is increased beyond 
about $0.01 M_{gal}$.
In fact the situation is even more dramatic than this, as 
illustrated in Figure 5, which shows the behavior of boxlike 
orbits integrated for just 10 orbital periods in triaxial models 
with a smooth core and various values of $M_{BH}/M_{gal}$.
When the black hole mass exceeds $\sim 2\%$ of the galaxy mass, 
orbital evolution takes place in only a few oscillations --
just the behavior expected in the globally-stochastic regime.
If one imagines slowly increasing the mass of a black hole 
at the center of a triaxial galaxy, Figure 5 suggests that 
the galaxy would be forced to respond very rapidly -- perhaps 
in just a few crossing times -- once the black hole mass exceeded 
$\sim0.02\ M_{gal}$.

\section Self-Consistent Evolution

In recent studies of the triaxial self-consistency 
problem$^{16,17,27}$, attempts were made to identify the stochastic orbits 
and to treat them differently from the regular orbits in the model 
construction.
The separation of regular from stochastic orbits in these 
studies was based on the detection of linear instability of 
the motion; regular orbits were defined simply as those that 
showed no clear evidence of instability over $\sim 100$ 
oscillations.
But the true situation is more complex, as discussed above.
Boxlike orbits exhibit a wide range of diffusion rates, with 
no clear separation into ``regular'' and ``stochastic'' families.
In addition, orbital periods decrease toward the center of 
a galaxy; an orbit with a diffusion timescale of $10^3$  
periods would mimic a regular orbit in the outer parts of a 
galaxy, but would behave chaotically over a galaxy lifetime if 
located near the center.

There would seem to be no substitute for $N$-body codes when 
dealing with a situation as complex as this.
Until recently, $N$-body algorithms were unable to deal in a 
practical way with the high degree of central concentration 
of realistic triaxial models.
But the situation has changed, and a number of 
algorithms are now available that can efficiently represent the 
gravitational potentials of systems with central cusps 
and nuclear black holes.$^{28}$
Here, the results of a new $N$-body study$^{19}$ of triaxial 
galaxies with central singularities is presented.

The initial conditions for this study consisted of $\sim 10^5$ 
particles distributed in a nonrotating triaxial bar, with 
$c/a\approx 0.5$ and $b/a\approx 0.75$.
The initial model was generated by collapsing a cold, spherical 
cloud and allowing it to evolve to an equilibrium state; the triaxial 
shape resulted from a bar-making instability associated with cold 
collapse.$^{29}$
In order to increase the resolution very near the center, 
particles were given masses that depended on their initial 
positions; masses varied by a factor of 10 from the center to the envelope.
Particle positions were advanced with individual timesteps, sometimes 
differing by as much as a factor of $10^6$, using a
fourth-order integrator.

\figureps[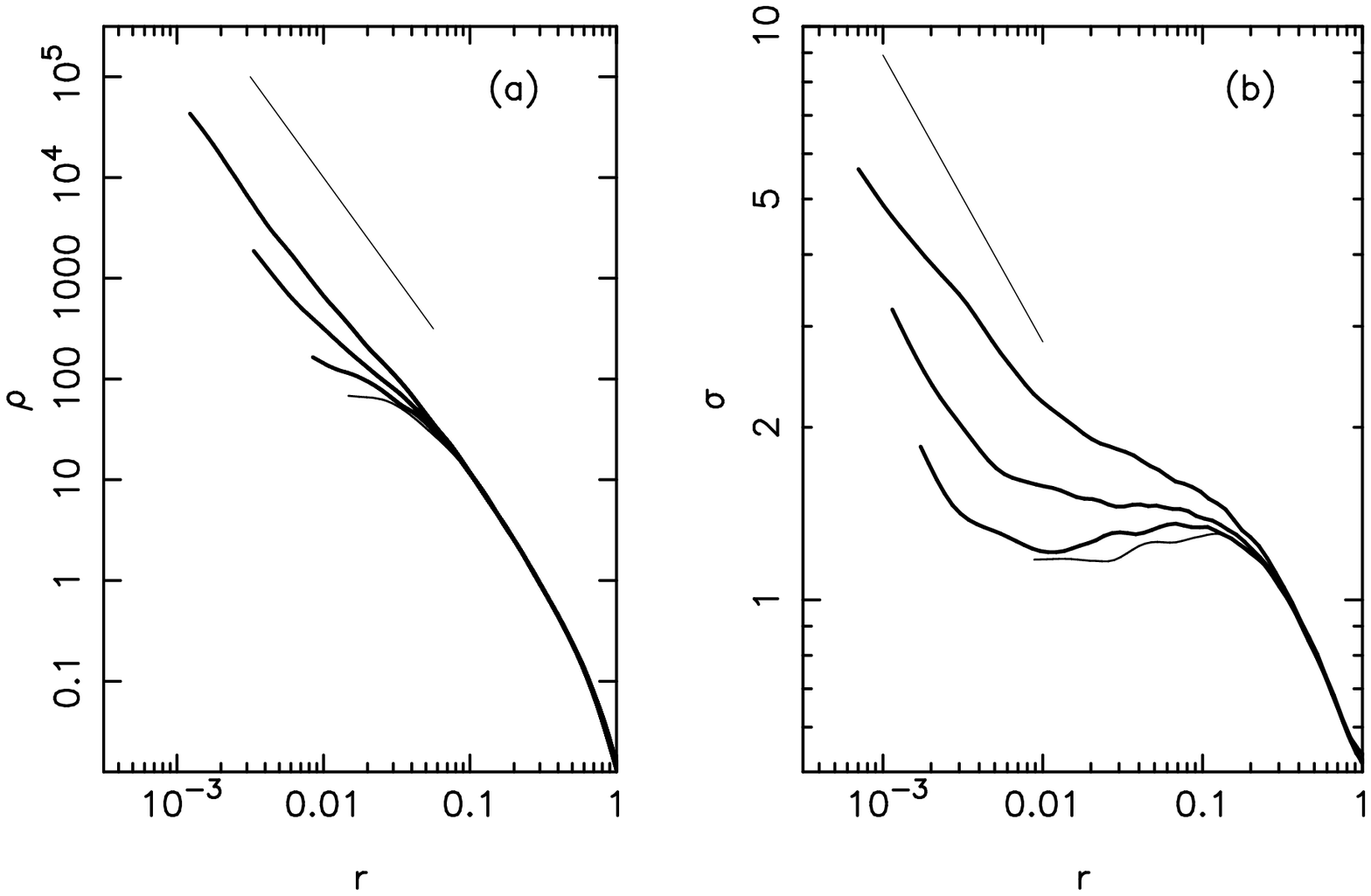,1.\hsize]
6. Growth of central density cusps in $N$-body 
models with various black hole masses.$^{19}$ (a) Density profiles; 
(b) velocity dispersion profiles.
The thin curve is the initial model without a black hole; the 
three heavy curves are final models with 
$M_{BH}/M_{gal}=0.003,0.01,0.03$.
The straight line in panel (a) has a logarithmic slope of $-2$; in panel 
(b), of $-0.5$.

The initial model had a constant-density core.
A ``black hole'' was grown in this core by increasing the mass 
at the origin according to $M(t)=M_{BH}\tau^2(3-2\tau)$, with 
$\tau=t/t_{grow}$.
Various values for $M_{BH}$ and $t_{grow}$ were used.
Figure 6 illustrates the formation of a density cusp in the stars 
surrounding the black hole; the cusp forms because the black hole 
pulls in the neighboring stars as the force of its attraction 
increases.$^{30}$

The growth of a black hole in an initially triaxial model is 
expected to make the model more axisymmetric, through a two-step 
process.$^{31, 32}$
Many of the box orbits in the initial model are rendered 
stochastic by the black hole; as a result, they evolve to fill a 
more-or-less spherical volume corresponding roughly to the region 
enclosed by an equipotential surface.
Nearly-spherical orbits are not very useful for reconstructing a 
barlike shape, and so the model responds by becoming more 
axisymmetric.
As it approaches axisymmetry, the tube orbits that circulate 
about the symmetry axis are able to reproduce the mass 
distribution self-consistently and the galaxy settles rapidly 
into equilibrium.

This picture appears to be essentially correct, as illustrated in 
Figure 7, which shows the evolution of the intermediate-to-short 
axis ratio $b/a$ of the model as the black hole is grown.
For three different black hole masses, $M_{BH}/M_{gal}=0.3\%, 1\%$ 
and $3\%$, the model evolves to a final state that is almost 
precisely axisymmetric.
The short-to-long axis ratio also increases, from its inital value 
of $\sim 0.5$, to $\sim 0.9$ near the center and $\sim 0.6$ 
at the half-mass radius; the elongation of the model ceases to 
change once axisymmetry is reached, a fact which argues in favor of the 
evolution being driven by the box orbits.

Although the final shape of the model is nearly the same for the 
three different values of $M_{BH}$ tested here, the evolution time 
was found to depend strongly on the black hole mass.
When $M_{BH}=0.003M_{gal}$, axisymmetry was not quite reached by the end of 
the integration, at roughly 55 half-mass orbital periods.
But when $M_{BH}$ was increased to $0.03M_{gal}$, the model 
evolved in shape 
on approximately the same timescale that the black hole grew.
Experiments with smaller values of $t_{grow}$ revealed that the 
response time of the galaxy to the black hole was essentially 
instantaneous -- i.e., of order the local orbital period -- when 
$M_{BH}$ exceeded $\sim2\%$ of the galaxy mass.
Such rapid evolution is just what would be expected based on the 
test particle integrations shown in Figure 5.
We conjecture that a transition to global stochasticity occurs at
$M_{BH}\approx 0.02\ M_{gal}$, causing the boxlike orbits to lose
their characteristic shapes in a single orbital period.
The galaxy responds by rapidly becoming axisymmetric.
 
\figureps[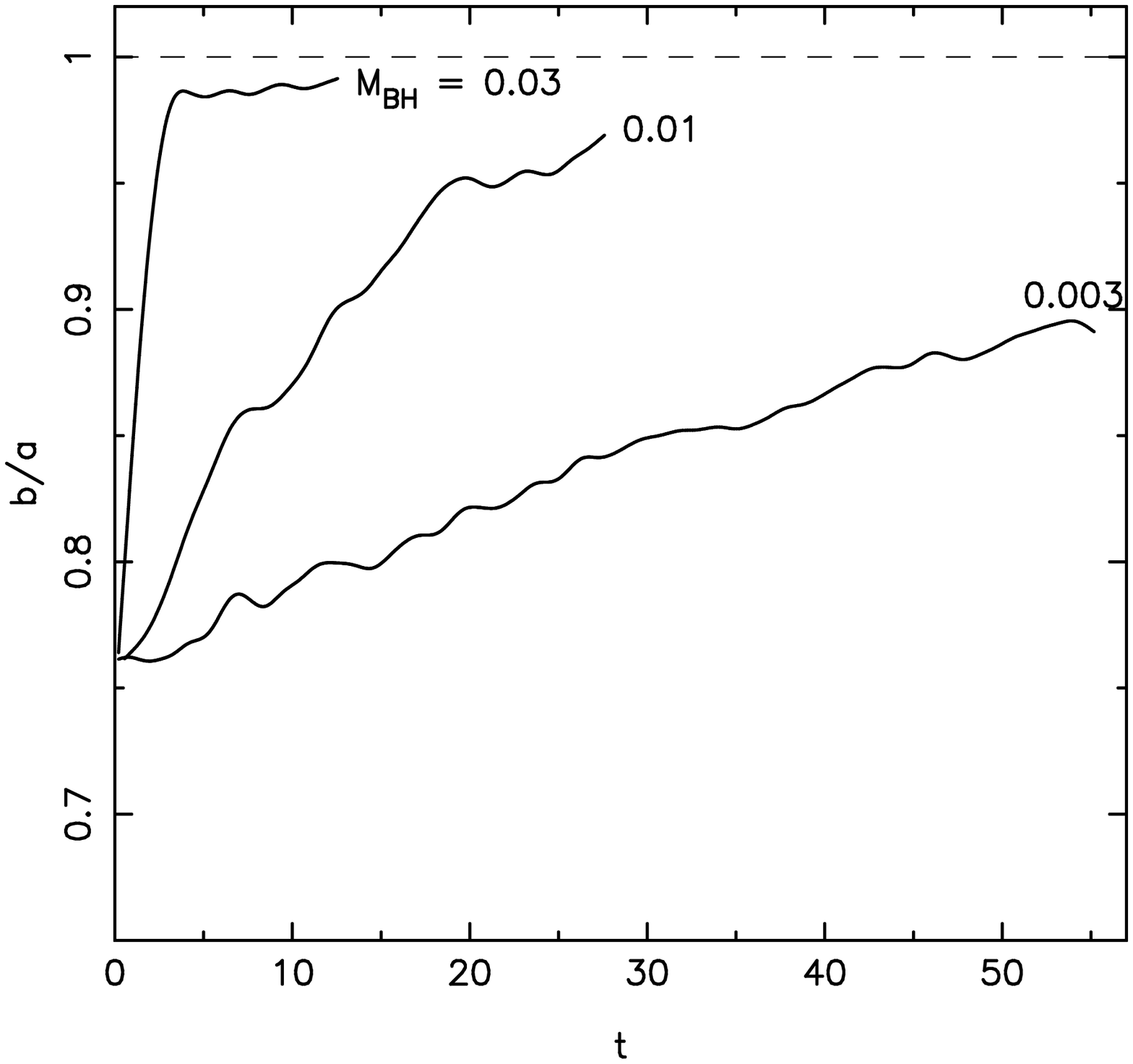,0.8\hsize]
7. Evolution of the intermediate-to-long axis ratio 
$b/a$ of the $N$-body models as nuclear black holes of three 
different masses are grown.$^{19}$ 
Time is in units of the half-mass orbital period.
Black hole masses are expressed in terms of the total mass of the 
galaxy model.
The axis ratio was computed from the most-bound 50\% of the 
stars.
The two smaller black holes were grown on a timescale of 5 half-mass 
orbital periods; the black hole with $M_{BH}=0.03$ was grown in 2 
half-mass orbital periods.

This rapid evolution toward axisymmetry may constitute a 
negative feedback mechanism that determines the maximum mass of 
nuclear black holes.$^{19}$
In one widely-discussed scheme,$^{33}$ nuclear black holes grow 
by capturing stars or gas clouds on box orbits as they pass near 
the galactic center.
But as the mass of the black hole approaches $\sim 2\%$ of the 
galaxy's mass, the galaxy would evolve rapidly to axisymmetry, 
thus eliminating the box orbits and cutting off the supply of 
stars or gas to the black hole.
Black holes could then never accrete more than $\sim 2\%$ of 
their host galaxy's mass unless they somehow managed to grow on 
a timescale much shorter than a galaxy crossing time.
This hypothesis is consistent with what is known so far about 
the masses of nuclear black holes, the largest of which 
have masses of order $1\%$ the stellar mass of their parent galaxies.$^8$
The current record-holder is NGC 3115,$^{34}$ which has 
$M_{BH}/M_{gal}\approx 2.5\%$, nicely consistent with the 
prediction.

Although the critical black hole mass found here was derived from 
a single triaxial model with a particular shape, roughly the same mass may 
define the transition to global stochasticity in models with
very different geometries. 
For instance, Norman, Sellwood \& Hasan $^{35}$ found that 2-D 
models of rotating barred spiral galaxies exhibited strong evolution 
toward axisymmetry when $M_{BH}/M_{gal}$ exceeded $\sim 5\%$. 
But in 3-D simulations of similar barred systems,$^{36}$ 
the bar was found to be much more fragile, evolving strongly once
the black hole mass fraction exceeded $\sim 1\%$.

\bigskip

\acknowl 
The $N$-body work described here was carried out in collaboration 
with G. Quinlan. 
We thank J. Sellwood for helpful comments on the manuscript.
This work was supported by NASA grant NAG 5-2803 and by NSF grants 
AST 93-18617 and AST 96-17088.

\references

\ 1. Chandrasekhar, S. 1942. Principles of Stellar Dynamics. 
University of Chicago Press. Chicago, IL.

\ 2. Hunter, C. 1995. Ann. NY Acad. Sci. {\bf 751}: 76

\ 3. Kuzmin, G. G. 1956. Astr. Zh. {\bf 33}: 27

\ 4. Kuzmin, G. G. 1973. Quadratic Integrals of Motion and 
Stellar Orbits in the Absence of Axial Symmetry of the Potential. 
{\it In} Dynamics of Galaxies and Clusters. T. B. Omarov, Ed.: 
71. Akad. Nauk. Kaz. SSR. Alma Ata.

\ 5. de Zeeuw, P. T. 1985. Mon. Not. R. Astron. Soc. {\bf 216}: 
273.

\ 6. Gebhardt et al. 1996. Astron. J. {\bf 112}: 105.

\ 7. Merritt, D. \& T. Fridman. 1995. Equilibrium 
and Stability of Elliptical Galaxies. {\it In} Fresh Views of 
Elliptical Galaxies, A. S. P. Conf. Ser. Vol. {\bf 86}. A. 
Buzzoni, A. Renzine \& A. Serrano, Eds.: 13-22.
Astronomical Society of the Pacific. Provo, Utah.

\ 8. Kormendy, J. \& D.  O. Richstone. 1995. Ann. Rev. Astron. 
Astrophys. {\bf 33}: 581.

\ 9. Goodman, J. \& M. Schwarzschild. 1981. Astrophys. J. {\bf 
245}: 1087.

\ 10. Fridman, T. \& D. Merritt. 1997. Astron. J. {\bf 114}: in 
press.

\ 11. Miralda-Escud\'e, J. \& M. Schwarzschild. 1989. 
Astrophys. J. {\bf 339}: 752.

\ 12. Gerhard, O. \& J. J. Binney. 1985. Mon. Not. R. 
Astron. Soc. {\bf 216}: 467.

\ 13. Schwarzschild, M. 1979. Astrophys. J. {\bf 232}: 236.

\ 14. Schwarzschild, M. 1982. Astrophys. J. {\bf 263}: 599.

\ 15. Statler, T. 1987. Astrophys. J. {\bf 321}: 113.

\ 16. Merritt, D. \& T. Fridman. 1996. Astrophys. J. {\bf 460}: 
136. 

\ 17. Merritt, D. 1997. Astrophys. J. {\bf 486}: in press.

\ 18. Valluri, M. \& D. Merritt. 1997. Rutgers Astrophysics 
Preprint Series No. 214.

\ 19. Merritt, D. \& G. Quinlan. 1997. Rutgers Astrophysics 
Preprint Series No. 212.

\ 20. Merritt, D. \& M. Valluri. 1996. Astrophys. J. {\bf 471}: 
82.

\ 21. Laskar, J., C. Froeschl\'e \& A. Celletti. 1992. Physica 
D {\bf 56}: 253.

\ 22. Laskar, J. 1996. Introduction to Frequency Map Analysis.
{\it In} NATO-Advanced Study Institute, Hamiltonian Systems with 
Three or More Degrees of Freedom. C. Simo \& A. Delshams, Eds.

\ 23. Papaphilippou, Y. \& J. Laskar. 1996. Astron. Astrophys.
{\bf 307}: 427.

\ 24. Papaphilippou, Y. \& J. Laskar. 1997. Astron. Astrophys.
in press.

\ 25. Dehnen, W. 1993. Mon. Not. Royal Astron. Soc. {\bf 265}: 
250.

\ 26. Laskar, J. 1993. Physica D {\bf 67}: 257.

\ 27. Schwarzschild, M. 1993. Astrophys. J. {\bf 409}: 563

\ 28. Sellwood, J. A. 1997. Galaxy Dynamics by $N$-Body 
Simulation. {\it In} Computational Astrophysics, A. S. P. Conf. Ser. 
Vol. {\bf 123}. D. A. Clarke \& M. J. West, Eds.: 215-220.
Astronomical Society of the Pacific. Provo, Utah.

\ 29. Aguilar, L. \& D. Merritt. 1990. Astrophys. J. {\bf 354}: 
33.
 
\ 30. Peebles, P. J. E. 1972. Gen. Rel. Grav. {\bf 3}: 63. 

\ 31. Gerhard, O. \& J. Binney. 1995. Mon. Not. Royal Astron. 
Soc. {\bf 216}: 467.

\ 32. Norman, C. A., A. May \& T. S. van Albada. 1985. 
Astrophys. J. {\bf 296}: 20.

\ 33. Norman, C. A. \& J. Silk. 1993. Astrophys. J. {\bf 266}: 
502.

\ 34. Kormendy, J., R. Bender \& S. Tremaine. 1996. Astrophys. 
J. {\bf 459}: L57. 

\ 35. Norman, C. A., J. A. Sellwood \& H. Hasan. 1996. 
Astrophys. J. {\bf 462}: 114.

\ 36. Sellwood, J. A. \& E. Moore, in preparation.

\bye